\documentclass[
    ,final            
  ]
  {aipproc}

\layoutstyle{6x9}

\begin{document}

\title{Jet modification in hot and dense QCD matter}


\author{Guang-You Qin} {address={Department of Physics, Duke University, Durham, NC 27708, USA}}

\begin{abstract}

High energy quarks and gluons play essential roles in the tomographic study of relativistic heavy-ion collisions and of the quark-gluon plasma.
They interact with the traversed dense matter through elastic and inelastic collisions, and lose some of their initial energy in the process.
Experimental observables for jet modification include single inclusive hadron production, hadron emission opposite to a triggered hadron or photon, and full jet observables.
Here a brief review of parton energy loss and jet quenching in high energy nucleus-nucleus collisions is presented.

\end{abstract}

\maketitle


\section{Introduction}

The main goal of relativistic heavy-ion programs performed at RHIC and the LHC is to produce and study a new state of matter, the quark-gluon plasma (QGP).
Large transverse momentum partonic jets are regarded as useful tools for exploring various properties, such as energy density and temperature, of such highly excited dense matter \cite{Majumder:2010qh}.
During their propagation through the QGP, they are modified by interacting with the constituents of the traversed matter.
Jet modification and jet energy loss have been confirmed by multiple experimental observations, one of which is the significant suppression of single inclusive hadron  production at high transverse momenta \cite{Adcox:2001jp, Adler:2002xw}.

Jet modification is a combinational effect of drag, diffusion and stimulated emission experienced by the partonic jet shower when propagating through the dense nuclear medium.
Various transport coefficients, e.g., $\hat{e}=dE/dt$, $\hat{e}_2 = d(\Delta E)^2/dt$ and $\hat{q} = d(\Delta p_T)^2/dt$ are used to quantify the medium modification of jet shower \cite{Baier:1996kr, Qin:2012cz}.
There currently exist a number of perturbative QCD (pQCD) based schemes for the treatment of the radiative part of parton energy loss (for a systematic comparison, please refer to Ref. \cite{Armesto:2011ht}).
Various phenomenological calculations have been performed to describe the observed jet quenching data, such as single inclusive hadron suppression \cite{Wicks:2005gt, Qin:2007rn, Armesto:2009zi, Chen:2010te}, and di-hadron correlations \cite{Zhang:2007ja, Renk:2008xq} and photon-hadron correlations \cite{Qin:2009bk, Zhang:2009rn}.

In recent years, sophisticated experimental techniques have been employed to obtain the fully reconstructed jets in high energy nucleus-nucleus collisions.
This provides a more stringent test on our understanding of jet energy loss in terms of the medium modification of both leading and subleading fragments of the jet shower \cite{Vitev:2009rd}.
The study of full jet evolution helps to understand how jets deposit energy into medium and how the medium responds to the deposited energy \cite{Qin:2009uh}.
The heavy-ion program at the LHC has extended the collision energy by more than a factor of ten compared to RHIC, opening up a
new era of jet quenching study in relativistic nuclear collisions.

Here a short review is presented for the current status on our theoretical understanding of jet modification and parton energy loss in dense nuclear matter within the framework of pQCD, and the progress on the phenomenological studies of hadron and jet production at large transverse momenta at RHIC and the LHC.

\section{Single inclusive hadron production}

The suppression of single inclusive hadron spectra at large transverse momenta can be quantified by the nuclear modification factor $R_{AA}$,
\begin{eqnarray}
R_{AA} = \frac{1}{N_{\rm coll}} \frac{dN_h^{AA}/d^2p_T dy}{dN_h^{pp}/d^2p_T dy}
\end{eqnarray}
The production of high transverse momentum hadrons in elementary proton-proton collisions can be well described by pQCD.
Schematically, it can be computed through the following formula within the factorization paradigm of QCD,
\begin{eqnarray}
d\sigma_h = f_a \otimes f_b \otimes d\sigma_{ab \to jd} \otimes D_{j\to h}
\end{eqnarray}
where $\sigma_{ab \to jd}$ is the short distance parton-parton cross section, $f_a$, $f_b$ are parton distribution functions, and $D_{j \to h}$ is the fragmentation function.
In relativistic nucleus-nucleus collisions, the partonic jet $j$ produced from initial hard parton-parton collisions will interact with the dense matter before fragmenting into hadrons. One may include the medium effect by replacing the vacuum fragmentation function $D_{j \to h}$ with a medium-modified fragmentation function $\tilde{D}_{j \to h} = P_{j \to j'} \otimes D_{j'\to h}$,
where the function $P_{j \to j'}$ encodes the effect of medium modification on jet propagation. It may depends on jet energy or virtuality, and also on detailed information of the medium that the jet traversed, such as the evolution profiles of medium energy density, temperature and flow. Relativistic hydrodynamics have been very successful in describing the evolution of the soft bulk matter which serves as the background for jet quenching study.

One of the most systematic jet quenching studies was performed in Ref. \cite{Bass:2008rv} where three different jet energy loss models 
were used to compute single hadron observables at high transverse momenta within the same hydrodynamical background. The main finding from this study is that while different models present very similar description of jet quenching data at RHIC, the jet transport parameters, e.g., $\hat{q}$, differ by a factor of two up to five. This result has stimulated a lot of theoretical studies to identify the origin of model differences. In Ref. \cite{Armesto:2011ht}, currently available pQCD based formalisms for radiative energy loss
are systematically compared for the QGP ``brick" problem. The differences between energy loss models persist for the brick problem, and the largest difference can be attributed to the specific approximations such as eikonal and soft collinear approximations that different models made when calculating the spectrum of induced gluon radiation.
This study has clearly revealed the deficiencies of the current parton energy loss formalisms, and provided the strong motivation to develop full next-to-leading order (NLO) jet quenching schemes with the incorporation of both soft collinear and hard large angle radiations and match between them.

The measurement of single inclusive hadron $R_{AA}$ at the LHC is one of the most exciting results for jet quenching study \cite{Aamodt:2010jd, CMS:2012aa}.
Predictions for $R_{AA}$ based on pQCD theory of parton energy loss are qualitatively consistent with the data \cite{Muller:2012zq}.
The rapid rise of $R_{AA}$ as a function of transverse momenta is attributed to a much flatter spectrum for the leading parton at the LHC compared to RHIC.
Meanwhile, large qualitative differences are found among different energy loss models and the data.
This may imply that some important effects that have been neglected in previous model implementations should be included, such as the running coupling \cite{Betz:2012qq}, virtualities \cite{Muller:2010pm}, finite size effect \cite{CaronHuot:2010bp}, etc.

\section{Full jet observables}

The application of sophisticated experimental techniques to fully reconstruct jets enable us to obtain detailed information including both leading and subleading fragments of a jet shower.
The much higher energy of jets available at the LHC allows the investigation of medium effects on jets with transverse energies in excess of 100~GeV.
One of the first exciting full jet measurements from the LHC is the energy asymmetry $A_J$ between correlated jet pairs in Pb+Pb collisions \cite{Aad:2010bu, Chatrchyan:2011sx},
\begin{eqnarray}
\label{eq:AJ}
A_J = \frac{p_{T,1} - p_{T,2}}{p_{T,1} + p_{T,2}} ,
\end{eqnarray}
where $p_{T,i}, (i=1,2)$ denote transverse momenta of the leading and sub-leading jets for a chosen jet size $R$, respectively.
While the modification of dijet energy imbalance increases from peripheral to central Pb+Pb collisions, no strong change is observed for the distribution of the dijet azimuth separation $\Delta \phi = |\phi_1 - \phi_2|$.
This indicates that the subleading jets may experience significant energy loss during propagating through the hot and dense medium created in Pb+Pb collisions.

Various theoretical models have been employed to explain the observed dijets asymmetry in Pb+Pb collisions at the LHC \cite{CasalderreySolana:2010eh, Qin:2010mn, Lokhtin:2011qq, Young:2011qx, He:2011pd, ColemanSmith:2011rw, Renk:2012cx}.
To understand the energy loss from the jet cone, one needs to calculate not only the energy loss from the primary parton of the jet, but the evolution of shower gluons in the medium as well.
The in-medium evolution of shower gluons may be described by the following equation \cite{Qin:2010mn},
\begin{equation}
\label{eq:dG/dt}
\frac{d}{dt}f_g(\omega, k_\perp^2, t) = \hat{e} \frac{\partial f_g}{\partial \omega}
  + \frac{1}{4} \hat{q} {\nabla_{k_\perp}^2 f_g} +  \frac{dN_g^{\rm med}}{d\omega dk_\perp^2 dt} .
\end{equation}
The first and second terms describe elastic energy loss and transverse momentum broadening experienced by the shower gluons, and the third describes the gluon radiation of the propagating jet.
Combined with the evolution of leading parton and radiated shower gluons, one may obtain the energy dissipated from the jet cone.
Calculations have shown that the observed dijet asymmetry in Pb+Pb collisions may be explained by the effect that a lot of low energy gluons are removed from the jet cone by the medium, either by deflection or by energy dissipation and thermalization \cite{CasalderreySolana:2010eh, Qin:2010mn}.
One not yet fully answered question is how the lost energy from the jet cone is thermalized, dissipated and redistributed in the hydrodynamic background medium. This requires one to evolve the jet shower in medium and the hydrodynamic response to jet transport simultaneously.

\section{Electromagnetic probes}

Additional information about jet-medium interaction and jet energy loss may be obtained from electromagnetic radiation in relativistic heavy-ion collisions.
Theoretical calculations indicated that there might be sizable contribution from jet-medium interaction to the photon production at intermediate transverse momenta \cite{Qin:2009bk}. On the other hand, photons at high transverse momenta serve as the baseline evidence for jet quenching as they are mainly produced from initial hard collisions and not modified in nucleus-nucleus collisions compared to proton-proton collisions.

High transverse momentum photons can be used as a tag for studying jet quenching; one looks at the medium modification of away-side correlated jets.
Such a channel has been regarded as the ``golden" channel for jet quenching study since it provides more constrained information about the away-side tagged jets at the production point \cite{Wang:1996yh}. Also many of the biases present in studying medium modification for single inclusive jets or dijets are removed, such as the deep falling spectra, surface bias, tangential emission, etc.
The measurements of photon-triggered fragmentation functions at RHIC \cite{Adare:2009vd, Abelev:2009gu} and the modification of energy imbalance between correlated photon-jet pairs at the LHC \cite{Chatrchyan:2012gt} are qualitatively consistent with the picture of jet energy loss and redistribution of lost energy.
More details about photon-tagged jets, such as their longitudinal and transverse profiles, would provide more discriminative power against various parton energy loss models.

\section{Summary}

Jet quenching is one of the most important tools to study the transport properties of hot and dense nuclear medium in relativistic heavy-ion collisions.
It is a remarkable success that the leading parton energy loss in a strongly interacting quark-gluon plasma is consistent with pQCD based energy loss models.
The fully reconstructed jet observables and correlation measurements have provided a lot more detailed information about medium-modified jets and put more stringent constraints on modeling jet modification and energy loss in dense matter.

In the mean time, various uncertainties and deficiencies still exist in and between different parton energy loss models, mainly originating from the soft, collinear and eikonal approximations made in the calculation of radiated gluon spectrum, and the treatments of multiple gluon emissions.
There are yet still many questions not fully answered.
What are the relative contributions from radiative and collisional components of jet energy loss?
What do the extracted values of jet transport coefficients $\hat{e}$ and $\hat{q}$ tell us about the detailed medium structure that jets probe?
How does the lost energy dissipate and thermalize in the medium, and affect the reconstruction of jets?
As compared to hadronic observables, how much additional constraint can we gain from the fully reconstructed jets \cite{Renk:2012cx}?

In summary, to obtain a full understanding of jet shower modification in dense nuclear matter, it is required to extend the current energy loss theories and models to incorporate some important missing ingredients, such as full NLO calculation of medium-induced gluon radiation and consistent treatment of collisional and radiative components of energy loss.
Jet quenching should become more quantitative in describing various experimental observed quantities such as single inclusive hadron production, full jet observables as well as correlation measurements.
The large kinematics spanning from RHIC to LHC provides a crucial tool for discriminating different jet energy loss models.
The ultimate goal is to not only obtain a unified description of jet modification in medium as well as the medium response to jet transport, but to gain deep insights of the structure of the quark-gluon plasma created in relativistic heavy-ion collisions.




\begin{theacknowledgments}

This work was supported in part by the U. S. Department of Energy under grants DE-FG02-05ER41367 and (within the framework of the JET Collaboration) de-sc0005396.

\end{theacknowledgments}



\bibliographystyle{aipproc}   




\bibliographystyle{h-physrev5}
\bibliography{GYQ_refs}

\begin{thebibliography}{10}

\bibitem{Majumder:2010qh}
A.~Majumder and M.~Van~Leeuwen,
\newblock Prog.Part.Nucl.Phys. {\bf A66}, 41 (2011).

\bibitem{Adcox:2001jp}
PHENIX, K.~Adcox {\em et~al.},
\newblock Phys. Rev. Lett. {\bf 88}, 022301 (2002).

\bibitem{Adler:2002xw}
STAR, C.~Adler {\em et~al.},
\newblock Phys. Rev. Lett. {\bf 89}, 202301 (2002).

\bibitem{Baier:1996kr}
R.~Baier, Y.~L. Dokshitzer, A.~H. Mueller, S.~Peigne, and D.~Schiff,
\newblock Nucl. Phys. {\bf B483}, 291 (1997).

\bibitem{Qin:2012cz}
G.-Y. Qin and A.~Majumder,
\newblock (2012), arXiv:1205.5741.

\bibitem{Armesto:2011ht}
N.~Armesto {\em et~al.},
\newblock (2011), arXiv:1106.1106.

\bibitem{Wicks:2005gt}
S.~Wicks, W.~Horowitz, M.~Djordjevic, and M.~Gyulassy,
\newblock Nucl. Phys. {\bf A784}, 426 (2007).

\bibitem{Qin:2007rn}
G.-Y. Qin {\em et~al.},
\newblock Phys. Rev. Lett. {\bf 100}, 072301 (2008).

\bibitem{Armesto:2009zi}
N.~Armesto, M.~Cacciari, T.~Hirano, J.~L. Nagle, and C.~A. Salgado,
\newblock J. Phys. {\bf G37}, 025104 (2010).

\bibitem{Chen:2010te}
X.-F. Chen, C.~Greiner, E.~Wang, X.-N. Wang, and Z.~Xu,
\newblock Phys. Rev. {\bf C81}, 064908 (2010).

\bibitem{Zhang:2007ja}
H.~Zhang, J.~F. Owens, E.~Wang, and X.-N. Wang,
\newblock Phys. Rev. Lett. {\bf 98}, 212301 (2007).

\bibitem{Renk:2008xq}
T.~Renk,
\newblock Phys. Rev. {\bf C78}, 034904 (2008).

\bibitem{Qin:2009bk}
G.-Y. Qin, J.~Ruppert, C.~Gale, S.~Jeon, and G.~D. Moore,
\newblock Phys. Rev. {\bf C80}, 054909 (2009).

\bibitem{Zhang:2009rn}
H.~Zhang, J.~Owens, E.~Wang, and X.-N. Wang,
\newblock Phys.Rev.Lett. {\bf 103}, 032302 (2009).

\bibitem{Vitev:2009rd}
I.~Vitev and B.-W. Zhang,
\newblock Phys. Rev. Lett. {\bf 104}, 132001 (2010).

\bibitem{Qin:2009uh}
G.~Y. Qin, A.~Majumder, H.~Song, and U.~Heinz,
\newblock Phys. Rev. Lett. {\bf 103}, 152303 (2009).

\bibitem{Bass:2008rv}
S.~A. Bass {\em et~al.},
\newblock Phys. Rev. {\bf C79}, 024901 (2009).

\bibitem{Aamodt:2010jd}
ALICE Collaboration, K.~Aamodt {\em et~al.},
\newblock Phys.Lett. {\bf B696}, 30 (2011).

\bibitem{CMS:2012aa}
CMS Collaboration, S.~Chatrchyan {\em et~al.},
\newblock Eur.Phys.J. {\bf C72}, 1945 (2012).

\bibitem{Muller:2012zq}
B.~Muller, J.~Schukraft, and B.~Wyslouch,
\newblock (2012), arXiv:1202.3233.

\bibitem{Betz:2012qq}
B.~Betz and M.~Gyulassy,
\newblock Phys.Rev. {\bf C86}, 024903 (2012).

\bibitem{Muller:2010pm}
B.~Muller,
\newblock Nucl.Phys. {\bf A855}, 74 (2011).

\bibitem{CaronHuot:2010bp}
S.~Caron-Huot and C.~Gale,
\newblock Phys.Rev. {\bf C82}, 064902 (2010).

\bibitem{Aad:2010bu}
Atlas Collaboration, G.~Aad {\em et~al.},
\newblock Phys.Rev.Lett. {\bf 105}, 252303 (2010).

\bibitem{Chatrchyan:2011sx}
CMS Collaboration, S.~Chatrchyan {\em et~al.},
\newblock Phys.Rev. {\bf C84}, 024906 (2011).

\bibitem{CasalderreySolana:2010eh}
J.~Casalderrey-Solana, J.~G. Milhano, and U.~A. Wiedemann,
\newblock J.Phys.G {\bf G38}, 035006 (2011).

\bibitem{Qin:2010mn}
G.-Y. Qin and B.~Muller,
\newblock Phys.Rev.Lett. {\bf 106}, 162302 (2011).

\bibitem{Lokhtin:2011qq}
I.~Lokhtin, A.~Belyaev, and A.~Snigirev,
\newblock Eur.Phys.J. {\bf C71}, 1650 (2011).

\bibitem{Young:2011qx}
C.~Young, B.~Schenke, S.~Jeon, and C.~Gale,
\newblock Phys.Rev. {\bf C84}, 024907 (2011).

\bibitem{He:2011pd}
Y.~He, I.~Vitev, and B.-W. Zhang,
\newblock Phys.Lett. {\bf B713}, 224 (2012).

\bibitem{ColemanSmith:2011rw}
C.~Coleman-Smith, G.-Y. Qin, S.~Bass, and B.~Muller,
\newblock AIP Conf.Proc. {\bf 1441}, 892 (2012).

\bibitem{Renk:2012cx}
T.~Renk,
\newblock Phys.Rev. {\bf C85}, 064908 (2012).

\bibitem{Wang:1996yh}
X.-N. Wang, Z.~Huang, and I.~Sarcevic,
\newblock Phys.Rev.Lett. {\bf 77}, 231 (1996).

\bibitem{Adare:2009vd}
PHENIX, A.~Adare {\em et~al.},
\newblock Phys. Rev. {\bf C80}, 024908 (2009).

\bibitem{Abelev:2009gu}
STAR, B.~I. Abelev {\em et~al.},
\newblock Phys. Rev. {\bf C82}, 034909 (2010).

\bibitem{Chatrchyan:2012gt}
CMS Collaboration, S.~Chatrchyan {\em et~al.},
\newblock (2012), arXiv:1205.0206.

\end{thebibliography}


\end{document}